\documentstyle[twocolumn,twoside,prb,floats,aps]{revtex}
\input{psfig}
\tighten 
\begin{document}
\draft

\title{Doubled shot noise in disordered 
normal-metal--superconductor junctions}

\author{M. J. M. de Jong$^{a,b}$ and C. W. J. Beenakker$^{b}$}

\address{ 
(a) Philips Research Laboratories,
5656 AA  Eindhoven,
The Netherlands\\
(b) Instituut-Lorentz,
University of Leiden,
2300 RA  Leiden,
The Netherlands\\
{\rm (Submitted 28 February 1994)}\\
\parbox{14 cm}{\medskip\rm\indent
The low-frequency shot-noise power of 
a normal-metal--superconductor junction is studied for arbitrary
normal region. Through a scattering approach, a
formula is derived which expresses the shot-noise
power in terms of the transmission eigenvalues of the normal region.
The noise power divided by the current is enhanced
by a factor two with respect to its normal-state value, due to
Cooper-pair transport in the superconductor.
For a disordered normal region, it is still
smaller than the Poisson noise, as a consequence of noiseless open
scattering channels.}}
\maketitle
\narrowtext

Electrical shot noise is the time-dependent fluctuation of the current around
the average $I$, due to the discreteness of the charge carriers.
The shot-noise power $P$ gives information on the conduction process
which is not contained in the resistance.
A well-known example is a vacuum diode, where
$P=2 e |I|
\equiv P_{\text{Poisson}}$. This tells
us that the electrons traverse the
diode in completely uncorrelated fashion, as in a Poisson process.
A noise power of $P_{\text{Poisson}}$ is the maximum 
value in the normal state (N). In macroscopic samples shot noise is fully
suppressed due to inelastic processes. For samples of dimensions smaller
than the inelastic scattering length shot noise is observable, but may be
suppressed below $P_{\text{Poisson}}$ due to correlated electron
transmission \cite{l&b91}.
In this paper we  investigate the enhancement of shot noise in
disordered normal-metal--superconductor (NS) junctions.
Naively, one would expect $P=4 e |I|= 2 P_{\text{Poisson}}$,
since the current in the superconductor is carried by
Cooper pairs in units of $2e$.
Instead, we find $P=\frac{2}{3} P_{\text{Poisson}}$, due to noiseless
open scattering channels. We also consider the more general case of a 
disordered region in series with a tunnel barrier. In the absence of
disorder we recover previous results by Khlus \cite{khl87}. Independent
work on this problem has
been carried out by 
Musykantsky and Khmel'nitski\u{\i} \cite{mus94}.
We furthermore would like to mention recent work on shot noise
in a normal-metal--superconductor--normal-metal junction
\cite{han94}.

We first review the results for phase-coherent transport
in the normal state.
The conductance at zero temperature and small applied voltage $V$
is given by the Landauer formula
\begin{equation}
G_{\text{N}}=G_0 \mbox{Tr} \, {\bf t t}^\dagger =
G_0 \sum \limits_{n=1}^{N} T_n \: ,
\label{e1}
\end{equation}
where $G_0 \equiv 2 e^2/h$.
The matrix product ${\bf t t^\dagger}$ has eigenvalues $T_n$,
$n=1,2,\ldots N$, with $N$ the number of scattering channels
at the Fermi energy $E_F$ and $\bf t$ the transmission matrix.
 From current conservation it follows that $T_n \in [0,1]$.
A formula for the zero-frequency
shot-noise power has been derived by
B\"{u}ttiker \cite{but90},
\begin{equation}
P_{\text{N}} = P_0 \mbox{Tr} \, \left[ 
{\bf t t}^\dagger ( {\bf 1} - {\bf t t}^\dagger) \right]
= P_0 \sum \limits_{n=1}^{N} T_n (1 - T_n) \: ,
\label{e2}
\end{equation}
with $P_0 \equiv 2 e |V| G_0$.
Equation (\ref{e2}) is the multi-channel generalization of earlier
single-channel formulas \cite{khl87,les89}.
It is a consequence of the Pauli principle that
closed ($T_n=0$) as well as open ($T_n=1$) scattering channels
do not fluctuate and therefore give no contribution to the shot noise.

In the case of a tunnel barrier, all transmission eigenvalues
are small ($T_n \ll 1$, for all $n$), so that the quadratic terms in 
Eq.\ (\ref{e2}) can be neglected. Then it follows from comparison
with Eq.\ (\ref{e1}) 
that $P_{\text{N}}=2e|V|G_{\text{N}}=2 e |I|=P_{\text{Poisson}}$.
In contrast, for a quantum point contact $P_{\text{N}} \ll
P_{\text{Poisson}}$.
Since on the plateaus of quantized conductance
all the $T_n$'s are either 0 or 1, the shot noise is expected to be
only observable at the steps between the plateaus \cite{les89}. 
This is indeed confirmed in an experiment 
by Li {\em et al.} \cite{li90}.
For a diffusive conductor of length $L$ much longer than
the elastic mean free path $\ell$ it has been predicted that 
$P_{\text{N}}= \frac{1}{3} P_{\text{Poisson}}$, as a consequence of noiseless
open scattering channels
\cite{b&b92,nag92,jon92,naz94}. 
Recently, an experimental observation of suppressed shot noise in
a disordered wire has been reported \cite{lie94}.

\begin{figure}
\centerline{\psfig{file=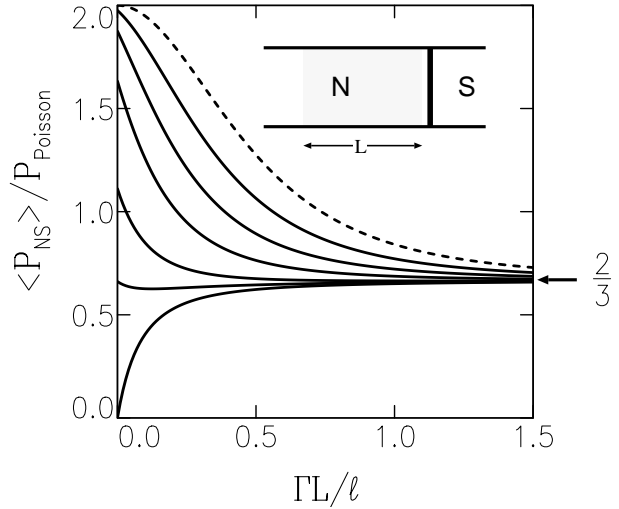,width=8.0cm}}
\vspace{0.5cm}
\caption{The shot-noise power of a NS junction 
(in units of $P_{\text{Poisson}} \equiv 2 e |I|$)
as a function of
the length $L$ (in units of $\ell/\Gamma$), for barrier transparencies 
$\Gamma= 1,0.9,0.8,0.6,0.4,0.2$
from bottom to top. The dashed curve gives the limiting result for
$\Gamma \ll 1$. For $L=0$ the noise power varies as a function
of $\Gamma$ according to Eq.\ (\protect\ref{e18}),
between doubled shot noise ($\langle P_{\text{NS}} \rangle=4e |I|$) 
for high barriers
($\Gamma \ll 1$) and zero in the absence of a barrier ($\Gamma=1$).
If $L$ increases the noise power approaches the limiting value
$\langle P_{\text{NS}} \rangle=\frac{4}{3} e |I|$ for 
{\em each} $\Gamma$.
The inset shows schematically the NS junction.
\label{f1}}
\end{figure}

Now, let us turn to transport through a NS junction.
The conducting properties have originally been described by
Blonder, Tinkham, and Klapwijk \cite{btk82}, and more recently in
Refs.\ \onlinecite{lam91,tak92,bee92}.
If the applied voltage
is smaller than the superconducting gap ($e|V| < \Delta$), the
dissipative normal current is converted at the NS interface into
dissipationless supercurrent,
by means of Andreev reflection: 
Electrons in the normal metal
are retro-reflected at the NS interface into holes, with the 
transfer of a Cooper pair to the superconducting condensate. 
The scattering geometry is illustrated in the inset to Fig.\ \ref{f1}.
Electrons and holes, incident from a reservoir via 
an ideal (impurity-free) lead, are scattered by an arbitrarily
disordered, normal region in series with a superconductor.
The applied voltage is taken to be small, and the temperature low,
so that transmission of excitations into the superconductor is
prohibited. All incident quasiparticles are therefore reflected back
into the reservoir.

The calculation of the shot-noise power of the
NS junction proceeds along the lines of B\"{u}ttiker's method
for  normal-metal conductors \cite{but90}.
In the present case the scattering states 
are solutions of the Bogoliubov-de Gennes 
equation \cite{btk82,lam91,tak92,bee92}, 
rather than of a single-particle Schr\"{o}dinger equation.
The current operator in the lead towards the NS junction is
given by 
\begin{equation}
\hat{I}(t) = \frac{e}{h} \sum_{\alpha, \beta}
\int \limits_0^\infty d\varepsilon
\int \limits_0^\infty d\varepsilon'
I_{\alpha \beta}(\varepsilon, \varepsilon')
\hat{a}_\alpha^\dagger(\varepsilon)
\hat{a}_\beta(\varepsilon')
e^{it(\varepsilon - \varepsilon')/\hbar} \; ,
\label{e4}
\end{equation}
where $\hat{a}_\alpha^\dagger(\varepsilon)$ 
[$\hat{a}_\alpha(\varepsilon)$] is the creation (annihilation)
operator of scattering state $\psi_\alpha(\varepsilon)$,
and $I_{\alpha \beta}(\varepsilon, \varepsilon')$ is the matrix element of
the current operator between states $\psi_\alpha(\varepsilon)$ and 
$\psi_\beta(\varepsilon')$.
The quasiparticle
energy $\varepsilon$  is measured with respect to $E_F$.  
In the lead, the  state $\psi_\alpha$ consists of one
incoming mode $\varphi_\alpha^+$ and several, reflected, outgoing modes
$\varphi_\beta^-$,
\begin{equation}
\psi_\alpha(\varepsilon) =
\varphi_\alpha^+(\varepsilon) + \sum \limits_\beta
r_{\beta \alpha}(\varepsilon) \varphi_\beta^-(\varepsilon)
\; .
\label{e5}
\end{equation}
The indices $\alpha, \beta$ denote
mode number $(m)$ as well as whether it concerns electron
[$\alpha=(m,e)$] or hole [$\alpha=(m,h)$] propagation.
The modes $\varphi^+, \varphi^-$
are normalized to carry unit quasiparticle flux.
The reflection amplitudes $r_{\beta \alpha}$ are contained in the
unitary $2N \times 2 N$ matrix ${\bf r}$, which has the block form
\begin{equation}
{\bf r} = 
\left( \begin{array}{ll}
{\bf r}_{\vphantom{h}ee} & {\bf r}_{eh} \\
{\bf r}_{he} & {\bf r}_{hh} 
\end{array} \right)
\; ,
\label{e6}
\end{equation}
where e.g.\ the $N \times N$ submatrix
${\bf r}_{he}$ contains the reflection amplitudes from incoming
electrons to reflected holes.
The unitarity of the reflection matrix corresponds to 
conservation of the number of quasiparticles.
The conductance of the NS junction is given by
\cite{tak92}
\begin{equation}
G_{\text{NS}} = 2 G_0  \mbox{Tr} \, 
{\bf r}^{\vphantom{\dagger}}_{he} {\bf r}_{he}^\dagger \; .
\label{e12}
\end{equation}

In the zero-frequency limit we need the current-matrix elements
$I_{\alpha \beta}(\varepsilon, \varepsilon)$ at equal energies.
Following Ref.\ \onlinecite{but90}, we find
\begin{equation}
I_{\alpha \beta}(\varepsilon, \varepsilon) = 
\left[{\bf \Lambda} -  {\bf r}^\dagger(\varepsilon)  {\bf \Lambda}
{\bf r}(\varepsilon)\right]_{\alpha \beta} \; .
\label{e8}
\end{equation}
The difference with Ref.\ \onlinecite{but90} is the inclusion of 
the $2N \times 2N$ matrix ${\bf \Lambda}$, defined by
\begin{equation}
{\bf \Lambda} \equiv \left( \begin{array}{cc}
-{\bf 1} & {\bf 0} \\
{\bf 0} &  {\bf 1}
\end{array} \right) \; ,
\label{e9}
\end{equation}
which accounts for the opposite charges of electrons and holes.
The average current $I$ can be determined
from the expectation value
of Eq.\ (\ref{e4}), using
\begin{equation}
\langle 
\hat{a}_\alpha^\dagger(\varepsilon)
\hat{a}_\beta(\varepsilon')
\rangle = \delta_{\alpha \beta} \delta( \varepsilon - \varepsilon')
f_\alpha(\varepsilon) \; ,
\label{e10}
\end{equation}
with $f_\alpha(\varepsilon)$ the distribution function in the reservoir.
At zero temperature and for $V<0$ one has 
for the electron ($f_e$) and hole ($f_h$) distribution
functions
\begin{equation}
f_e(\varepsilon) = \Theta( e |V| - \varepsilon) \; , \hspace{1cm}
f_h(\varepsilon) = 0 \; ,
\label{e11}
\end{equation}
with $\Theta(x)$ the unit-step function.
The conductance 
$G_{\text{NS}}\equiv\lim_{V \rightarrow 0} I/V$
can now easily be determined from
Eqs.\ (\ref{e4}), (\ref{e8}),
(\ref{e10}), and (\ref{e11}). This indeed
provides the result Eq.\ (\ref{e12}) of Ref.\ \onlinecite{tak92},
which serves as a check on the formalism.

We are now ready to compute
the zero-frequency shot-noise power, defined by
\begin{equation}
P_{\text{NS}} \equiv 2 \int \limits_{-\infty}^{\infty} dt  
\left\langle \Delta \hat{I}(t) \Delta \hat{I}(0) \right\rangle \; ,
\label{e13}
\end{equation}
with $\Delta \hat{I}(t) \equiv \hat{I}(t) - I$.
Substituting Eq.\ (\ref{e4}) and using Eq.\ (\ref{e10})
we find
\begin{equation}
P_{\text{NS}} = 2 \frac{e^2}{h} \int \limits_0^\infty d\varepsilon
\sum \limits_{\alpha, \beta} 
I_{\alpha \beta}(\varepsilon, \varepsilon)
I_{\beta \alpha} (\varepsilon, \varepsilon)
f_\alpha(\varepsilon) \left[ 1 - f_\beta(\varepsilon ) \right]
\; .
\label{e14}
\end{equation}
Equation (\ref{e14}) can be evaluated through 
Eqs.\ (\ref{e8}) and (\ref{e11}). 
In the zero-temperature, zero-voltage limit we find, 
making use of the unitarity of $\bf r$,
\begin{eqnarray}
P_{\text{NS}} &=& 4 P_0  \mbox{Tr} 
\left[ {\bf r}_{he}^{\vphantom{\dagger}} {\bf r}_{he}^\dagger 
({\bf 1} -  {\bf r}_{he}^{\vphantom{\dagger}} {\bf r}_{he}^\dagger )  \right]
\nonumber \\
&=& 4 P_0 \sum \limits_{n=1}^N {\cal R}_n ( 1 - {\cal R}_n )
 \; ,
\label{e16}
\end{eqnarray}
where ${\cal R}_n$ is an eigenvalue of 
${\bf r}_{he}^{\vphantom{\dagger}} {\bf r}_{he}^\dagger$,
evaluated at $\varepsilon=0$.

It remains to relate the Andreev-reflection eigenvalues ${\cal R}_n$ to
the scattering properties of the
normal region. 
In the presence of time-reversal symmetry, i.e.\ in zero magnetic
field, the eigenvalues ${\cal R}_n$ can be expressed entirely in terms of
the transmission eigenvalues $T_n$ of the {\em normal} region \cite{bee92} :
\begin{equation}
{\cal R}_n = T_n^2 (2 - T_n)^{-2} \; .
\label{e16a}
\end{equation}
Equation (\ref{e16a}) assumes a step function (at the NS interface) for the
pair potential and neglects terms of order
$(\Delta/E_F)^2$. 
Substitution into Eq.\ (\ref{e12}) yields the result of 
Ref.\ \onlinecite{bee92}
for the conductance of the NS junction,
\begin{equation}
G_{\text{NS}}= G_0 \sum \limits_{n=1}^N \frac{2 T_n^2}{(2 - T_n)^2 } \; .
\label{e3}
\end{equation}
We now apply the same method to our result (\ref{e16}) for
the shot-noise power, and find 
\begin{equation}
P_{\text{NS}} = P_0 \sum \limits_{n=1}^N 
\frac{16 T_n^2 (1 - T_n)}{(2 - T_n)^4} \; .
\label{e17}
\end{equation}
This is our main result. 
It is a general formula for arbitrary disorder potential in the normal region.
As in the normal state, scattering channels which have 
$T_n=0$ or $T_n=1$ do not contribute to the shot noise.
However, the way in which partially transmitting channels contribute
is entirely different from the normal state result (\ref{e2}).
Before considering the case of a disordered conductor, we first briefly
discuss the case of a planar tunneling barrier, which was previously
studied by Khlus \cite{khl87}.

A planar tunnel barrier is modeled by a channel-independent
barrier transparency: $T_n = \Gamma$,
for all $n$.  It follows from Eq.\ (\ref{e2}), 
that for a normal conductor this would yield
$P_{\text{N}}=(1 - \Gamma) P_{\text{Poisson}} $, implying full 
Poisson noise for
a high barrier ($\Gamma \ll 1$). For the NS junction we find from 
Eqs.\ (\ref{e3}) and (\ref{e17})
\begin{equation}
P_{\text{NS}} = P_0 N \frac{16 \Gamma^2 (1 - \Gamma)}{(2 - \Gamma)^4} =
\frac{8 (1 - \Gamma)}{(2 - \Gamma)^2} P_{\text{Poisson}} 
\; .
\label{e18}
\end{equation}
This agrees with the result of Khlus \cite{khl87,not2}.
If $\Gamma < 2(\sqrt{2}-1) \approx 0.83$, 
one observes a shot noise {\em above}
the Poisson noise.
For $\Gamma \ll 1$ one has 
\begin{equation}
P_{\text{NS}}= 4 e |I| = 2 P_{\text{Poisson}} \; , 
\label{e19}
\end{equation}
which is a {\em doubling} of the shot-noise power divided by the current with
respect to the normal-state result. 
This can be interpreted as an uncorrelated
current of $2e$-charged particles.

We now turn to a NS junction with a disordered normal region, 
of length $L$ much greater than
the mean free path $\ell$, but much smaller than the localization length,
so that transport is in the metallic, diffusive regime.  
In Ref.\ \onlinecite{b&b92}  the average of the
normal-state shot-noise power is computed.
The method  is applicable to
any physical quantity of the form $\sum_n f(T_n)$ with
$\lim_{T \rightarrow 0} f(T)=0$. (Such a quantity is called a linear
statistic on the transmission eigenvalues.)
Our formula (\ref{e17}) for the shot noise in the NS junction
is of this form. According to Ref.\ \onlinecite{b&b92} one has the general
formula
\begin{equation}
\left\langle \sum \limits_{n=1}^{N} f(T_n) \right\rangle =
\left\langle \sum \limits_{n=1}^{N} T_n \right\rangle
\int \limits_0^\infty dx \, f( \cosh^{-2} x ) \; .
\label{e19a}
\end{equation}
Equation (\ref{e19a}) is obtained from the
relationship $T_n=\cosh^{-2}(L /\zeta_n)$
between the transmission eigenvalues and the 
channel-dependent localization lengths $\zeta_n$, and from  
the fact that $L/\zeta$ is uniformly
distributed between 0 and $L/\ell \gg 1$.
This uniform distribution is a
general result of random-matrix theory \cite{stoMP}, but
has also been derived from a microscopic Green's function theory \cite{naz94}.
The ensemble-averaged shot-noise power is now easily calculated
by application of Eq.\ (\ref{e19a}) to Eqs.\ (\ref{e3}) and (\ref{e17}),
with the result 
\begin{equation}
\frac{\left\langle P_{\text{NS}} \right\rangle}
{\left\langle G_{\text{NS}} \right\rangle} = 
\frac{2}{3} \, \frac{P_0}{G_0}
\; ,
\label{e20}
\end{equation}
hence
\begin{equation}
\left\langle P_{\text{NS}} \right\rangle=
\frac{4}{3} e |I| = \frac{2}{3} P_{\text{Poisson}}
\; .
\label{e20a}
\end{equation}
Equation (\ref{e20a}) is twice the result in the
normal state, but still smaller than the
Poisson noise.
Corrections to (\ref{e20a}) are of lower order in $N$ and due
to quantum-interference effects \cite{jon92}. 

Finally, we discuss a normal region which contains a disordered part as
well as a tunnel barrier. This is most relevant to
experiments, because in practice the NS interface is almost never ideal,
but has a transparency $\Gamma < 1$. 
However, the uniform distribution of $L/\zeta$ does not apply
to such a system.
In Refs.\ \onlinecite{naz94}
and \onlinecite{bee94b} the distribution of transmission eigenvalues 
of such a system is studied and an
expression for $\langle G_{\text{NS}} \rangle$ 
as a function of $s\equiv L/\ell$ and $\Gamma$
is obtained. The shot-noise power can be derived in a similar fashion.
Here we merely present the final expressions,
\begin{mathletters}
\begin{eqnarray}
\left\langle G_{\text{NS}} \right\rangle &=& G_0 N \,
\frac{2 v'(\phi )}{2 s v'(\phi )-1}
\; , \label{e21a} \\
\left\langle P_{\text{NS}} \right\rangle &=& P_0 N \, \left(
\frac{4 v'(\phi )}{3 \left(2 s v'(\phi ) -1 \right)} - 
\frac{ 4 s {{v''(\phi )}^2}}{{{\left(2 s v'(\phi ) -1 \right) }^5}}
\right. \nonumber \\ 
&& \hphantom{ P_0 N } \left.  + \,
\frac{2 v'''(\phi )}{3 {{\left(2 s v'(\phi ) -1 \right) }^4}}
\right)
\; , \label{e21b}
\end{eqnarray}
\end{mathletters}%
with $v'(\phi ), \: v''(\phi ), \: v'''(\phi )$ the first, second, and third
derivative of
\begin{equation}
v(\phi) \equiv 
\frac{\cos \phi }{2 / \Gamma + \sin \phi  -1 }
\; . \label{e22}
\end{equation}
The auxiliary variable $\phi \in (0,\pi/2)$ is the solution of
\begin{equation}
\phi = 2  s  v(\phi) \; .
\label{e23}
\end{equation}
The result is given in Fig.\ \ref{f1}, where
$\langle P_{\text{NS}} \rangle /P_{\text{Poisson}}$ is plotted against
$\Gamma L /\ell$ for various
$\Gamma$. Note, the crossover from the ballistic 
(\ref{e18}) to the diffusive result (\ref{e20a}).
For a high barrier ($\Gamma \ll 1$), the shot noise decreases
from twice the Poisson noise to two-thirds the Poisson noise 
as the amount of disorder increases. 

In summary, we have presented a theory for the shot noise in
normal--superconductor junctions for arbitrary normal region. 
The general result (\ref{e17}) can be applied to many mesoscopic systems.
We predict that for a disordered normal region
the shot noise is suppressed below the Poisson noise
by a factor $\frac{2}{3}$,  due
to the presence of noiseless open scattering channels.
This result is double the normal-state result,
a consequence of the Cooper-pair transport in the superconductor. 
For a normal region consisting of a disordered part and a
barrier (at the NS interface), the shot-noise power may vary between
zero and a doubled Poisson noise, depending on the junction
parameters.
We feel that observation of our predictions 
is within reach of present technology and presents a challenge
for experimentalists.  

This research was supported 
by the Dutch Science Foundation NWO/FOM.


\begin{references}
\bibitem{not2}
To see this,  one must evaluate Eq.\ (25a) of Ref.\ \onlinecite{khl87} for
$eV \ll \Delta$. Note, that the first exponent $2$ should be $-2$.
\end{references}
\end{document}